# Quantifying the factors limiting rate-performance in battery electrodes

Ruiyuan Tian,[1,2+] Sang-Hoon Park,[1,3+] Paul J. King,[4] Graeme Cunningham,[1,3] Joao Coelho,[1,3] Valeria Nicolosi,[1,3] Jonathan N Coleman[1,2*]

[1]*CRANN and AMBER research centers, Trinity College Dublin, Dublin 2, Ireland*

[2]*School of Physics, Trinity College Dublin, Dublin 2, Ireland.*

[3]*School of Chemistry, Trinity College Dublin, Dublin 2, Ireland*

[4]*Efficient Energy Transfer Department, Bell Labs Research, Nokia, Blanchardstown Business & Technology Park, Snugborough Road, Dublin 15, Ireland*

*colemaj@tcd.ie (Jonathan N. Coleman); Tel: +353 (0) 1 8963859.

[+]These authors contributed equally

ABSTRACT: A significant problem associated with batteries is the rapid reduction of charge-storage capacity with increasing charge/discharge rate. For example, improving this rate-performance is required for fast-charging of car batteries. Rate-performance is related to the timescales associated with charge or ionic motion in both electrode and electrolyte. However, no quantitative model exists which can be used to fit experimental data to give insights into the dominant rate-limiting processes in a given electrode-electrolyte system. Here we develop an equation which can be used to fit capacity *versus* rate data, outputting three parameters which fully describe rate-performance. Most important is the characteristic time associated with charge/discharge which can be expressed by a simple equation with terms describing each rate-limiting process, thus linking rate-performance to measureable physical parameters. We have fitted these equations to ~200 data sets from ~50 papers, finding exceptional agreement, and allowing parameters such as diffusion coefficients or electrolyte conductivities to be extracted. By estimating relevant physical parameters, it is possible to show which rate-limiting processes are dominant in a given situation, facilitating rational design and cell optimisation. In addition, this model predicts the upper speed limit for Li/Na ion batteries, in agreement with the fastest electrodes in the literature.

Rechargeable batteries that utilize Li-or Na-ion chemistry are becoming increasingly important for a number of technological applications, including electric vehicles (EVs), portable electronics, and grid-scale energy storage systems.[1,2] While great strides have been made toward both electrode design and the development of high capacity materials, high-rate (power) performance still needs to be significantly improved for a range of applications.[3] In particular, high rate-performance is critical to fulfil the demands of emerging applications such as rapid charging or high power delivery.[4]

The problem with rate performance in batteries is based on the fact that, above some threshold charge or discharge rate, $R_T$, the maximum achievable capacity begins to fall off with charge/discharge rate. Essentially, this limits the amount of energy a battery can deliver at high power, or store when charged rapidly. This is a significant problem and has led to a number of approaches targeting the electrode,[5-8] the electrolyte [9] and the separator[10] with the aim of increasing $R_T$ and reducing the rate of capacity falloff above $R_T$.

Much work has been done to identify the factors effecting high-rate capacity. It is known that rate performance can be improved by decreasing active particle size,[11-13] and electrode thickness,[14-17] or by increasing solid-state diffusivity,[11] conductor content,[7,16,18] electrode porosity[16,19] as well as by optimizing electrolyte concentration[14,16] and viscosity.[16]

Based on such information, it is generally accepted that a number of factors contribute to limiting rate performance: electronic transport in electrodes, ion transport both in bulk electrolyte and electrolyte-filled pores, solid-state diffusion of ions in the active materials and electrochemical reactions at the electrode/electrolyte interface.[12,20-22] One would expect that speeding up any of these processes would improve rate-performance.

However, the systematic, quantitative analysis of these factors, and how they depend on the physical or structural properties of the electrodes, separator and electrolyte is still incomplete. A number of papers have used modelling or numerical simulation to calculate, for example, the evolution of lithium concentration profiles in electrodes. Such work has certainly improved our understanding of the issues effecting rate performance.[14,22-24] However, what is lacking is a simple model that can quantify the relative importance of the factors effecting rate-performance in a given situation. Critically, it must be possible to use such a model to fit experimental data, to assess performance or gain mechanistic insights. While a small number of models exist which can be used to fit capacity *versus* rate data, all are limited in that they only describe a single rate limiting mechanism (e.g. diffusion in the electrolyte[24,25] or solid particles[2]) or only describe the high-rate region.[24,25]

Here we have developed a semi-empirical equation which accurately describes the rate dependence of electrode capacity in terms of electrode properties, *via* the characteristic time associated with charge/discharge. Importantly, we derive a simple expression for this characteristic time which includes the mechanistic factors described above. Together, these equations accurately describe a wide range of data extracted from the literature. By fitting experimental data, parameters such as electrolyte conductivity or diffusion coefficients can be extracted, while by plotting the equations using appropriate values of physical parameters, performance predictions can be made.

MODEL DEVELOPMENT

This work was inspired by recent work on rate-limitations in supercapacitors.[26,27] Although most supercapacitors are diffusion-limited,[28] they can also be electrically-limited, especially when electrodes are fabricated from low-conductivity materials or are used without current collectors.[26,27] For electrically-limited supercapacitor electrodes, the specific capacitance, $C_{SC}/M$, depends on the scan rate, $\nu$, *via*[26]

$$\frac{C_{SC}}{M} = C_{SC,M}\left[1-\frac{\nu\tau_{SC}}{\Delta V}\left(1-e^{-\Delta V/\nu\tau_{SC}}\right)\right] \tag{1}$$

where $C_{SC,M}$ is the intrinsic specific capacitance, $\Delta V$ is the voltage window and $\tau_{SC}$ is the RC time constant associated with charging/discharging the supercapacitor. Unlike diffusion-limited supercapacitors where the high-rate capacitance scales with $\nu^{-1/2}$, equation 1 predicts a high-rate scaling of $C_{SC} \propto \nu^{-1}$.[27] This equation has proven to be very useful for understanding the electrical behaviour of composite supercapacitor electrodes[27] and, with some empirical modifications, it also fits diffusion-limited supercapacitors.[27] With this in mind, we believe that this equation can be modified empirically to describe rate effects in battery electrodes.

An empirical generalisation of equation 1 to describe batteries could be obtained by replacing capacitance, $C_{SC}$, with capacity, $C$, and substituting $\nu/\Delta V$ by a fractional charge/discharge rate, $R$. This will result in an equation that gives capacity which is constant at low rate but scales as $R^{-1}$ at high rate. However, at high rate, diffusion-limited battery electrodes often display capacities which decay with rate as $R^{-1/2}$.[24] To facilitate this, and to allow for the possibility of alternative high-rate $R$-dependences, we modify the equation slightly so that at high rates, it is consistent with $C \propto R^{-n}$, where $n$ is a constant:

$$\frac{C}{M} = C_M \left[ 1 - (R\tau)^n \left( 1 - e^{-(R\tau)^{-n}} \right) \right] \tag{2}$$

Here $C/M$ is the measured specific capacity, $C_M$ is the intrinsic specific capacity and $\tau$ is the characteristic time associated with charge/discharge. Although this equation is semi-empirical, as we will show, it has the right form to describe rate-behaviour in batteries. In addition, the parameters, particularly $\tau$, are physically relevant.

To demonstrate that Equation 2 has the appropriate properties, in figure 1 we use it to generate plots of $C/M$ vs. $R$ for different values of $C_M$, $\tau$ and $n$. In all cases, we observe the characteristic plateau at low rate followed by a power-law decay at high rate. These graphs also make clear the role of $C_M$, $\tau$ and $n$. $C_M$ reflects the low-rate, intrinsic behaviour and is a measure of the maximum achievable charge storage. Taylor-expanding the exponential in equation 2 (retaining the first three terms) gives the high-rate behaviour:

$$\left( \frac{C}{M} \right)_{highR} \approx \frac{C_M}{2(R\tau)^n} \tag{3}$$

confirming a power-law decay with exponent $n$, a parameter which should depend on the rate-limiting mechanisms, with diffusion-limited electrodes displaying $n$=1/2. However, by analogy with supercapacitors, other values of $n$ may occur when electrodes are not solely diffusion limited (i.e. $n$=1 for resistance-limited behaviour).[26]

Most importantly, $\tau$ is a measure of $R_T$, the rate marking the transition from flat, low-rate behaviour to high-rate, power-law decay (transition occurs roughly at $R_T = (1/2)^{1/n} / \tau$). This means $\tau$ is the critical factor determining rate performance. As a result, we would expect $\tau$ to be related to intrinsic physical properties of the electrode/electrolyte system and reflect the physics of battery operation.

Before data can be fitted, the rate must be carefully defined. Most papers use specific current density, $I/M$, or the C-rate. However, after careful consideration, we decided to define rate as

$$R = \frac{I/M}{(C/M)_E} \tag{4}$$

where $(C/M)_E$ is the experimentally-measured specific capacity (at a given current). This contrasts with the usual definition of $\text{C-rate} = (I/M)/(C/M)_{Th}$ , where $(C/M)_{Th}$ is the theoretical specific capacity. We chose this definition because $1/R$ is then the measured charge/discharge time suggesting that $\tau$-values extracted from fits will have a physical significance.

## RESULTS AND DISCUSSION

We extracted capacity *versus* rate data from a large number of papers (>200 rate-dependent data sets from >50 publications), in all cases, converting current or C-rate to $R$. We divided the data into three cohorts: I, standard lithium ion electrodes;[7,16,17,29-46] II, standard sodium ion electrodes;[47-61] and III, data from studies which systematically varied the content of conductive additive.[7,18,19,60,62-68] Then, we fitted each capacity-rate data set to equation 2 (see figure 2A and SI for examples) finding very good agreement in all cases (~95% of fits yield $R^2$>0.99). From each fit, we extracted values for $C_M$, $n$ and $\tau$. Because of the broad spectrum of materials studied, the obtained values of $C_M$ spanned a wide range. As this paper is focused on rate effects, we will not discuss the extracted values of $C_M$, and will refer to them only when necessary.

Shown in figure 2B are the extracted values of $n$ and $\tau$ for cohorts I and II. It is clear from this panel that $n$ is not limited to values of 0.5, as would be expected for diffusion limited systems but varies from ~0.25 to 2.0. In addition, $\tau$ varies over a wide range from <1s to >1 h.

It is well-known that rate-performance tends to degrade as the electrode thickness (or mass loading) is increased.[17] Thus, $\tau$ should depend on the electrode thickness, $L_E$, which turns out to be the case (figure 2C). Interestingly, over the entire data set, $\tau$ scales roughly as $L_E^2$ (solid line). From this scaling, we define a parameter, $Q$, which we denote the transport coefficient: $Q = L_E^2/\tau$ , such that electrodes with higher $Q$ will have better rate-performance. The frequency of occurrence of $Q$ for the samples from cohorts I and II is plotted as a histogram in figure 2D. This shows a well-defined distribution with $Q$ varying from $10^{-9}$-$10^{-13}$ $m^2/s$. As we will show below, $Q$ is the natural parameter to describe rate performance in electrodes. In addition, we show that the upper end of the $Q$-distribution represents the ultimate speed limit in Li/Na-ion battery electrodes.

In diffusive systems, the length scale explored, $L$, is related to the time elapsed, $t$, by $L=\sqrt{Dt}$, where $D$ is the diffusion coefficient. Thus, at first glance, the $L_E^2$ scaling in figure 2C seems to suggest that battery electrodes are predominantly limited by diffusion of cations within the electrode, i.e. $Q=D=L_E^2/\tau$ such that $D$ varies between $10^{-9}$-$10^{-13}$ $m^2/s$. However, such a conclusion would be incorrect, as we will demonstrate. To see this, we first examine the exponent, $n$.

This parameter is plotted *versus* $L_E$ in figure 2E and displays only very weak thickness-dependence. More interesting is the histogram showing the frequency of occurrence of $n$-values in cohorts I and II (figure 2F). This clearly shows that most samples do not display $n$=0.5 as would be expected for purely diffusion-limited systems. In fact, we can identify weak peaks for $n$=0.5 and $n$=1 with most of the data lying in between. In supercapacitors, $n$=1 indicates electrical limitations.[26,27] If this also applies to batteries, figure 2 suggests most reported electrodes to be governed by a combination of diffusion and electrical limitations. Interestingly, a small number of data sets are consistent with $n$>1, indicating a rate-limiting mechanism which is even more severe than electrical limitations. We note that the highest values of $n$ are associated with Si-based electrodes where unwanted reactions such as alloying or Li-plating effects may affect lithium storage kinetics.[69]

*Varying conductive additive content*

The contribution of both diffusion and electrical limitations becomes clear by analysing cohort III of literature data (papers varying conducting additive content). Shown in figure 3A are specific capacity *versus* rate data for anodes of GaS nanosheets mixed with carbon nanotubes at different mass fractions, $M_f$ (ref[7]). A clear improvement in rate performance can be seen as $M_f$, and hence the electrode conductivity, increases, indicating changes in $\tau$ and $n$. We fitted data extracted from a number of papers[7,18,19,60,62-68] to equation 2 and plotted $\tau$ and $n$ *versus* $M_f$ in figures 3B and C. These data indicate a systematic drop in both $\tau$ and $n$ with increasing electrode conductivity.

Figure 3B shows $\tau$ to fall significantly with $M_f$ for all data sets with some samples showing a thousand-fold reduction. Such behaviour is clearly not consistent with diffusion effects being the sole rate-limiting phenomenon. We interpret the data as follows: at low $M_f$, the electrode conductivity is low and the rate performance is electrically-limited. As $M_f$ increases, so does the conductivity, reducing the electrical limitations and shifting the rate-limiting factor toward diffusion. This is consistent with the fact that, for a number of systems we see $\tau$ saturating at high $M_f$, indicating that rate-limitations associated with electron transport have been removed. We emphasise that it is the out-of-plane conductivity which is important in battery electrodes because it describes charge transport between current collector and ion storage sites.[27] This is an important distinction as nanostructured, composite electrodes can be highly anisotropic with out-of-plane conductivities much smaller[27] than the typically reported in-plane conductivities.[8,18]

Just as interesting is the data for *n versus* $M_f$, shown in figure 3C. For all data sets, *n* transitions from *n*~1 at very low $M_f$ to *n*~0.5, or even lower, at high $M_f$. This is consistent with *n*=1 representing resistance-limited and *n*=0.5 representing diffusion-limited behaviour as is the case for supercapacitors.[27] Because, electrodes become predominately diffusion-limited at high $M_f$, the values of *n* tend to be lower in cohort III compared to cohort I and II, especially at high $M_f$, as shown in figure 3D. Also, some data sets show *n*-values below 0.5; as low as 0.23. It is not yet clear if these low *n*-values are physically significant or just data-scatter.

*The relationship between* $\tau$ *and physical properties.*

This data strongly suggests most battery electrodes to be limited by a combination of resistance and diffusion limitations. This can be most easily modelled considering the characteristic time associated with charge/discharge, $\tau$. This parameter is a measure of $R_T$ so is likely to be controlled by physical properties. This data outlined above implies that $\tau$ has both resistance and diffusive contributions. In addition, we must include the effects of the kinetics of the electrochemical reaction at the electrode/electrolyte interface. This can be done *via* the characteristic time associated with the reaction, $t_c$, which can be calculated *via* the Butler-Volmer equation,[20] and can range from ~0.1 to >100 s.[20]

Then, $\tau$ is the sum of the three contributing factors:

$$\tau = \tau_{Electrical} + \tau_{Diffusive} + t_c \qquad (5A)$$

It is likely that the diffusive component is just the sum of diffusion times associated with *cation* transport in the electrolyte, both within the separator (coefficient $D_S$) and the electrolyte-filled pores within the electrode (coefficient $D_P$) as well as in the solid active material (coefficient $D_{AM}$).[20] These times can be estimated using $L = \sqrt{Dt}$ such that

$$\tau_{Diffusive} = \frac{L_E^2}{D_P} + \frac{L_S^2}{D_S} + \frac{L_{AM}^2}{D_{AM}} \quad \text{(5B)}$$

where $L_E$, $L_S$ and $L_{AM}$ are the electrode thickness, separator thickness and the length-scale associated with active material particles, respectively. $L_{AM}$ depends on material geometry: for a thin film of active material, $L_{AM}$ is the film thickness while for a quasi-spherical particle of radius $r$,[20] $L_{AM}=r/3$.

For the electrical contribution, we note that every battery electrode has an associated *capacitance*[70] that limits the rate at which the electrode can be charged/discharged. This effective capacitance, $C_{eff}$, will be dominated by charge storage but may also have contributions due to surface or polarisation effects.[70] Then, we propose $\tau_{Electrical}$ to be the RC time constant associated with the circuit. The total resistance related to the charge/discharge process is the sum of the resistances due to out-of-plane electron transport in the electrode material ($R_{E,E}$), as well as ion transport, both in the electrolyte-filled pores of the electrode ($R_{I,P}$) and in the separator respectively ($R_{I,S}$). Then, the RC contribution to $\tau$ is given by

$$\tau_{Electrical} = C_{eff}(R_{E,E} + R_{I,P} + R_{I,S}) \quad \text{(5C)}$$

The overall characteristic time associated with charge/discharge is then the sum of capacitive, diffusive and kinetic components:

$$\tau = C_{eff}(R_{E,E} + R_{I,P} + R_{I,S}) + \frac{L_E^2}{D_P} + \frac{L_S^2}{D_S} + \frac{L_{AM}^2}{D_{AM}} + t_c \quad \text{(5D)}$$

We note that this approach is consistent with accepted concepts showing current in electrodes to be limited by both capacitive and diffusive effects.[71] The resistances in this equation can be rewritten in terms of the relevant conductivities ($\sigma$) using $R = L/(\sigma A)$, where $L$ and $A$ are the length and area of the region in question. In addition, both ion diffusion coefficients and conductivities in the pores of the electrode and separator can be related to their bulk-liquid values ($D_{BL}$ and $\sigma_{BL}$) *via* the porosity, $P$, using the Bruggeman equation,[72] ($D_{Porous} = D_{BL}P^{3/2}$ and $\sigma_{Porous} = \sigma_{BL}P^{3/2}$). As shown in the SI, this yields

$$\tau = L_E^2\left[\frac{C_{V,eff}}{2\sigma_E} + \frac{C_{V,eff}}{2\sigma_{BL}P_E^{3/2}} + \frac{1}{D_{BL}P_E^{3/2}}\right] + L_E\left[\frac{L_S C_{V,eff}}{\sigma_{BL}P_S^{3/2}}\right] + \left[\frac{L_S^2}{D_{BL}P_S^{3/2}} + \frac{L_{AM}^2}{D_{AM}} + t_c\right] \quad \text{(6a)}$$

Term 1 2 3 4 5 6 7

where $C_{V,eff}$ is the effective volumetric capacitance of the electrode (F/cm$^3$), $\sigma_E$ is the out-of-plane electrical conductivity of the electrode material, $P_E$ and $P_S$ are the porosities of the electrode and separator respectively. Here $\sigma_{BL}$ is the overall (anion and cation) conductivity of the bulk electrolyte (S/m). This equation has 7 terms which will refer to below as terms 1-7 (as labelled). Terms 1, 2 and 4 represent electrical limitations associated with electron transport in the electrode (1), ion transport in both the electrolyte-filled porous interior of the electrode (2) and separator (4). Terms 3, 5 and 6 represent diffusion limitations due to ion motion in the electrolyte-filled porous interior of the electrode (3) and separator (5) as well as solid diffusion within the active material (6). Term 7 is the characteristic time associated with the kinetics of the electrochemical reaction. We note that, as outlined below, for a given electrode, not all of these seven terms will be important. We can also write the equation with compound parameters, *a*, *b* and *c* to simplify discussion later:

$$\tau = aL_E^2 + bL_E + c \tag{6b}$$

If equation 6a is correct, then the falloff in $\tau$ with $M_f$ observed in figure 3B must be associated with term 1, *via* the dependence of $\sigma_E$ on $M_f$, which we can express using percolation theory:[27] $\sigma_E \approx \sigma_M + \sigma_0 (M_f)^s$, where $\sigma_M$ is the conductivity of the active material, and $\sigma_0$ and $s$ are constants (we approximate the conductivity-onset to occur at $M_f$=0 for simplicity). This allows us to write equation 6a as

$$\tau / L_E^2 \approx \frac{C_{V,eff} / 2}{\sigma_M + \sigma_0 (M_f)^s} + \beta_1 \tag{7}$$

where $\beta_1$ represents a combination of all other parameters. We extracted the most extensive data sets from figure 3B and reproduced them in figure 3E. We find very good fits, supporting the validity of equations 6a and 7. From the resultant fit parameters (see inset), we can work out the ratio of composite to matrix (i.e. active material) conductivities, $\sigma_E/\sigma_M$, which we plot *versus* $M_f$ in figure 3F. This shows that significant conductivity differences can exist between different conductive fillers, leading to different rate performances. As shown in the SI, by estimating $C_{V,eff}$, we can find approximate values of $\sigma_M$ and $\sigma_0$ which are in line with expectations.

*Thickness dependence*

Equation 6 implies a polynomial thickness dependence, rather than the $L_E^2$-dependence crudely suggested by figure 2C. To test this, we identified a number of papers which reported rate-dependence for different electrode thicknesses as well as preparing some electrodes and performing measurements ourselves. An example of such data is given in figure 4A for $LiFePO_4$-based lithium ion cathodes of different thicknesses,[17] with fits to equation 2 shown as solid lines. We fitted nine separate electrode thickness/rate-dependent data sets to equation 2 with the resultant $\tau$ and $n$ values plotted in figure 4B. Shown in figure 4C is $\tau$ plotted *versus* $L_E$ for each material with a well-defined thickness-dependence observed in each case. We fitted each curve to equation 6b, finding very good fits for all data sets, and yielding *a*, *b* and *c*.

We first consider the *c*-parameter (from equation 6, $c = L_S^2/(D_{BL}P_S^{3/2}) + L_{AM}^2/D_{AM} + t_c$). With the exception of μ-Si/NT ($c$=2027±264 s) and NMC/NT ($c$=3.6±1 s), the fits showed $c$~0 within error. This probably indicates the available $L_E$-ranges were too small to accurately obtain *c*. Because the 5$^{th}$ term in equation 6 is always small (typically $L_S$~25μm, $D_{BL}$~$3\times10^{-10}$ m$^2$/s and $P_S$~0.4, yielding ~1s) and assuming fast reaction kinetics (term 7), *c* is approximately given by $c \approx L_{AM}^2/D_{AM}$ and so is reflective of the contribution of solid-state diffusion to $\tau$ (term 6). Thus, the high values of *c* observed for the μ-Si samples are probably due to their large particle size (radius, *r*~0.5-1.5 μm measured by SEM). Combining the value of *c*=2027 s with reported diffusion coefficients for nano-Si ($D_{AM}$~$10^{-16}$ m$^2$/s),[73] and using the equation above with $L_{AM} = r/3$,[20] allows us to estimate $r = 3L_{AM} \approx 3\sqrt{cD_{AM}}$ ~1.3 μm, within the expected range.

That *c*~0 for most of the analysed data can be seen more clearly by plotting $\tau / L_E$ *versus* $L_E$ in figure 4D for a subset of the data (to avoid clutter). These data clearly follow straight lines with non-zero intercepts which is consistent with *c*=0 and *b*≠0 (from equation 6, $b = L_S C_{V,eff}/(\sigma_{BL}P_S^{3/2})$). The second point is important as it can only be the case in the presence of resistance limitations (the *b*-parameter is associated with resistance limitations due to ion transport in the separator).

We extracted the *a*- and *b*-parameters from the fits in figure 4C and plotted *a versus b* in figure 4E. The significance of this graph can be seen by noting that we can combine the definitions of *a* and *b* in equation 6 to eliminate $C_{V,eff}$, yielding

$$a = \left[\frac{\sigma_{BL}P_S^{3/2}}{\sigma_E} + \left(\frac{P_S}{P_E}\right)^{3/2}\right]\frac{b}{2L_S} + \frac{1}{D_{BL}P_E^{3/2}} \quad (8)$$

The value of $D_{BL}$ can be measured,[74] simulated[75] or simply estimated from the Stokes-Einstein equation[76] and tends to fall in a narrow range (1-5)×$10^{-10}$ $m^2$/s for common battery electrolytes.[74,75] Taking a midpoint of $D_{BL}$=3×$10^{-10}$ $m^2$/s and using $L_S$=25 μm (from the standard Celgard separator),[77] we plot equation 8 on figure 4E for two scenarios with extreme values of separator[78]/electrode porosity and different bulk-electrolyte to electrode conductivity ratios, first where $\sigma_{BL}/\sigma_E$ =3, $P_S$=0.5 and $P_E$=0.3 (blue) and also where $\sigma_{BL}/\sigma_E$ =0.05, $P_S$=0.3 and $P_E$=0.8 (red). We find the data to roughly lie in between these bounds. This shows the effect of electrode and separator porosities and identifies the typical range of $\sigma_{BL}/\sigma_E$ values. In addition, because electrolytes tend to have $\sigma_{BL}$ ~0.5 S/m,[79] this data implies the out-of-plane electrode conductivities to lie between 0.2 and 10 S/m for these samples. To test this, we measured the out-of-plane conductivity for one of our electrodes (SiGr/4%NT), obtaining 9.5 S/m, in good agreement with the model. Interestingly, the *a*-value for the GaS/NT electrodes of Zhang *et al.*[7] is quite large, suggesting a low out-of-plane conductivity. This is consistent with the NT $M_f$-dependence (figure 3F), taken from the same paper, which indicates relatively low conductivity-enhancement in this system.

From the definition of *b* (equation 6, $b = L_S C_{V,eff}/(\sigma_{BL} P_S^{3/2})$), we can estimate the effective volumetric capacitance, $C_{V,eff}$, for each material (estimating $\sigma_{BL}$ from the paper and assuming $P_S$=0.4 [ref[78]] and $L_S$=25 μm unless stated otherwise in the paper). Values of $C_{V,eff}$ vary in the range ~$10^3$-$10^5$ F/$cm^3$. To put this in context, typical commercial batteries have capacitances of ~1500 F (18650 cylindrical cell).[80] Assuming the electrodes act like series capacitors, gives a single electrode capacitance of ~3000 F. Approximating the single electrode volume as ~25% of the total yields an electrode volumetric capacitance of ~$10^3$ F/$cm^3$, similar to the lower end of our range. We found these $C_{V,eff}$ values to scale linearly with the intrinsic volumetric capacity of each material (=$\rho_E C_M$, where $\rho_E$ is the electrode density) as shown in figure 4F, indicating the capacitance to be dominated by charge storage effects. This slope is given by $C_{V,eff}/\rho_E C_M = 28$ F/mAh, a quantity which will prove useful for applying the model.

*Other tests of the characteristic time equation*

We can also test the veracity of equation 6 in other ways. The data of Yu *et al*[16] for electrodes with different conductivities, which was shown in figure 4C, has been replotted in figure 5A as $\tau/L_E$ *versus* $L_E$ and shows these composites to have roughly the same value of *b* (intercept) but significantly different values of *a* (slope). This is consistent with the electrode conductivity effecting term 1 in equation 6a, perfectly in line with the model.

We can also test the porosity-dependence predicted by equation 6, although care must be taken, as electrodes with varying porosity also tend to display varying conductivity, making it difficult to isolate the porosity-dependence. However Bauer *et al.*[19] describe rate performance of graphite/NMC electrodes with different porosities yet the same conductivity. Shown in figure 5B are $\tau / L_E^2$-values, found by fitting their data, plotted *versus* porosity. Equation 6 predicts that this data should follow

$$\frac{\tau}{L_E^2} = \left[ \frac{C_{V,eff}}{2\sigma_{BL}} + \frac{1}{D_{BL}} \right] P_E^{-3/2} + \beta_2 \qquad (9)$$

where $\beta_2$ is a combination of parameters. As there are only 2 data points, the curve of course matches the data. However, combining the fit parameters with estimates of $C_{V,eff}$ and $D_{BL}$ yields a value of $\sigma_{BL}$=0.5 S/m, in line with typical values of ~0.1-1 S/m.[79]

Yu *at al.*[16] reported rate-dependence for $LiFePO_4$ electrodes with various electrolyte concentrations, *c*. Shown in figure 5C are $\tau / L_E^2$-values, found by fitting their data, plotted *versus* 1/*c*. We can model this by replacing the electrolyte conductivity, $\sigma_{BL}$, in equation 6 using the Nearnst-Einstein equation, $\sigma_{BL} \approx F^2 c D_{BL} / t^+ RT$ as a reasonable approximation (here $t^+$ is the cation transport number which allows conversion between overall conductivity, $\sigma_{BL}$, and cation diffusion coefficient, $D_{BL}$, while the other parameters have their usual meaning). Then equation 6 predicts

$$\frac{\tau}{L_E^2} = \frac{t^+ RT}{F^2 D_{BL} c} \left[ \frac{C_{V,eff}}{2P_E^{3/2}} + \frac{L_S}{L_E} \frac{C_{V,eff}}{P_S^{3/2}} \right] + \beta_3 \qquad (10)$$

where $\beta_3$ is a combination of parameters. Fitting the data and estimating the various parameters as described in the SI allows us to extract $D_{BL} \approx 6.2 \times 10^{-11}$ m$^2$/s, close to the expected value of ~$10^{-10}$ m$^2$/s.

Another parameter which can be varied in principle but rarely in practice is the separator thickness ($L_S$). We varied this by using one, two and three stacked separators, measuring the rate performance of NMC/0.5%NT electrodes in each case. Values of $\tau / L_E^2$ extracted from the fits are plotted *versus* $L_S$ in figure 5D. Then equation 6 predicts

$$\frac{\tau}{L_E^2} = L_S\left[\frac{C_{V,eff}}{L_E\sigma_{BL}P_S^{3/2}}\right] + \beta_4 \qquad (11)$$

where $\beta_4$ is a compound parameter. Fitting the data and estimating parameters yields $\sigma_{BL}$~0.6 S/m, very similar to typical values of ~0.1-1 S/m.[79]

Equation 6 would imply the solid-state diffusion term (term 6) could be significant if $D_{AM}$ were small and $L_{AM}$ large, especially for low-$L_E$ electrodes. Ye *et al.*[12] measured rate dependence of electrodes consisting of thin nano-layers (<20 nm) of anatase $TiO_2$ deposited on highly-porous gold current collectors. In these systems, we expect solid-state diffusion to be limiting. The $\tau$-values found by fitting their data are plotted *versus* the $TiO_2$ thickness in figure 5E. Examining equation 6, we would expect this data to be described by

$$\tau = \frac{L_{AM}^2}{D_{AM}} + \beta_5 \qquad (12)$$

where $\beta_5$ is a compound parameter. This equation fits the data very well, yielding a solid-state diffusion coefficient of $D_{AM}$=3.3×10$^{-19}$ m$^2$/s, close to values of (2-6)×10$^{-19}$ m$^2$/s reported by Lindstrom *et al.*[81]

DISCUSSION

The data described above suggests that the combination of equations 2 and 6 can accurately describe capacity-rate data in battery electrodes. In practise, this model can be used in a number of ways, with the simplest being to fit experimental data. To do this, researchers would collect capacity vs. rate data for electrodes with different values of certain parameters such as electrode thickness, $L_E$, or conductor content, $M_F$, and use equation 6 to analyse the resultant $\tau$ data as shown in figures 4 or 5. Probably most useful is analysis of $\tau$ *versus* $L_E$, as it offers much information e.g. the importance of solid diffusion or reaction kinetics (from *c*), and an estimation of $C_{V,eff}$ (from *b*). We note the ability to fit data is a major advantage over more sophisticated models.

We can also use equation 6 to understand the balance of the different contributions to rate performance and so to design better electrodes. Earlier, we introduced the transport coefficient, *Q*, as a metric for rate performance. Applying equation 6a, we find an equation for *Q*:

$$Q^{-1} = \frac{\tau}{L_E^2} = \frac{C_{V,eff}}{2\sigma_E} + \frac{C_{V,eff}}{2\sigma_{BL} P_E^{3/2}} + \frac{1}{D_{BL} P_E^{3/2}} + \frac{C_{V,eff} L_S / L_E}{\sigma_{BL} P_S^{3/2}} + \frac{L_S^2 / L_E^2}{D_{BL} P_S^{3/2}} + \frac{L_{AM}^2 / L_E^2}{D_{AM}} + \frac{t_c}{L_E^2} \quad \text{(13a)}$$

Term: 1 2 3 4 5 6 7

Because $L_E$ has either been eliminated or mostly appears as a ratio with other lengths, this parameter is semi-intrinsic to the electrode/electrolyte system and the natural descriptor of rate performance. $Q$ is similar in form to a diffusion coefficient but describes diffusive, electrical and kinetic limitations. According to equation 2, rate performance will be maximised when both $n$ and $\tau$ are as small as possible. This allows us to consider $Q$ as a figure of merit for electrodes, with larger values of $Q$ indicating better rate performance. Thus, any strategy to improve rate performance must focus on maximising $Q$ (minimising $Q^{-1}$). Values of $Q$ can be put in context by figure 2D which show the practical upper limits to be $Q$~$10^{-9}$ $m^2/s$.

Writing equation 13a in this way allows another test of our model as it predicts $1/Q$ to be proportional to $C_{V,eff}$ . Because of the proportionality of $C_{V,eff}$ and the intrinsic volumetric capacity of the electrode ( $\rho_E C_M$ ), we can test this by plotting $1/Q$ *versus* $\rho_E C_M$ in figure 5F for cohorts I and II. We find a well-defined relationship, adding further support to our model. This graph is particularly important as it confirms the influence of $C_{V,eff}$ on electrode rate-performance. In addition, it highlights the unfortunate fact that high-performance electrode materials have an inherent disadvantage in terms of rate-behaviour.

It is important to realise what parameters can be controlled during any optimisation. $D_{BL}$ is limited by solvent effects, while $\sigma_{BL}$ is typically maximised at ~0.5 S/m.[79] $D_{AM}$ and $C_{V,eff}$ (via $\rho_E C_M$) are set by materials choice. $L_S$ and $P_S$ can in principle be varied but are limited by separator availability. While $L_E$ can be varied, enhancement of capacity will usually necessitate its maximisation. This means $\sigma_E$, $P_E$ and $L_{AM}$ are the only truly free parameters for optimisation.

Equation 13 also gives insight into parameter optimisation. All seven terms must be minimised for battery electrodes to display maximised rate performance (i.e. maximal $Q$). In figure 6, we have used equation 13a to plot the values of $Q^{-1}$ for each term as well as their sum *versus* five electrode parameters, $L_E$, $C_{V,eff}$, $\sigma_E$, $L_{AM}$ and $P_E$, using typical values for the remaining parameters. To avoid confusion, we plot $\tau$ (rather than $Q^{-1}$) *versus* $L_E$ in figure 6A. This shows solid diffusion to dominate thin electrodes (term 6) but electrical limitations associated with ions in electrode pores to be dominant for electrodes thicker then ~50 μm (term 2). In panels B-E, we plot $Q^{-1}$ as a function of each parameter. We find electrical limitations to be important for high-capacity electrode materials which also display high $C_{V,eff}$ (figure 6B). As is well-known, it is important to maximise the (out-of-plane) conductivity (figure 6C) to minimise its contribution to $Q^{-1}$. In thick electrodes, the effect of solid diffusion (figure 6D) is only important for the largest active-material particles. Interestingly, changing the electrode porosity (figure 6E) has a relatively small impact on $Q$. In addition, we note that term 5 is always relatively small and can be neglected in general. In addition, taking $t_c$=25 s, toward the upper end of the range given in ref[20], gives a reaction kinetics contribution (term 7) which is negligible compared to other terms under these circumstances (although reaction kinetics can be rate-limiting for thin electrodes.[82])

Equation 13a can be simplified considerably for electrodes with thickness >100 μm, as would be found in practical cells. Then, $Q$ is dominated by terms 2 and 4 with a non-negligible contribution from term 3 under certain circumstances. Specifically, because terms 1 and 2 scale in similar ways, term 1 can be ignored when it is much smaller than term 2 i.e. if $\sigma_E >> \sigma_{BL} P_E^{3/2}$. Taking $\sigma_{BL}$~0.5 S/m and $P_E$~0.5, this is true if $\sigma_E$ >>1 S/m, which should be the aim when introducing conductive additives. Term 6 can be completely neglected so long as it is smaller than term 3: $L_E / L_{AM} > \sqrt{D_{BL} P_E^{3/2} / D_{AM}}$. For $r$=60 nm ($L_{AM}$=20 nm) Si particles ($D_{AM}$~$10^{-16}$ $m^2/s$),[73] this is true if $L_E$>20 μm which will generally be the case in commercial electrodes. In addition, we neglect term 7 as $t_c / L_E^2$ should become relatively small for thick electrodes.

Under these circumstances, terms 1, 5, 6 and 7 in equation 13a are negligible, giving an approximate expression for $Q$. This equation can be generalised and simplified further by using the Nearnst-Einstein equation to eliminate $\sigma_{BL}$, allowing us to express $Q$ in terms of the electrolyte concentration, $c$:

$$Q \approx \frac{D_{BL} P_E^{3/2}}{1 + \frac{t^+ RT C_{V,eff}}{2F^2 c}\left(1 + 2\frac{L_S}{L_E}\left(\frac{P_E}{P_S}\right)^{3/2}\right)} \qquad (13b)$$

Inspection of equation 13b shows the maximum possible value of $Q$ is achieved when $C_{V,eff}$ is small and the electrode is limited solely by diffusion of ions in the electrolyte-filled pores of the electrode: $Q_{max} \approx D_{BL} P_E^{3/2}$, which could reach ~$3\times10^{-10}$ $m^2/s$ in high porosity electrodes. Because cation diffusion can never be eliminated in practical (i.e. high $L_E$) electrodes, $Q_{max}$ represents the basic rate-limit for the electrode and is indicated on figure 2D by the arrow.

Virtually all of the electrodes analysed in this work show $Q<Q_{max}$. Interestingly, a recent paper used complex fabrication techniques to prepare nanostructured electrodes with the aim of achieving ultrafast charge/discharge.[21] This work achieved very impressive rate performance. Fitting their data yielded a value of $Q\sim 3\times 10^{-10}$ m$^2$/s, very close to the maximum value suggested by our work.

We can use equation 13b to calculate how significantly $Q$ can fall below its maximum value. The only parameters in equation 13b which vary significantly in real systems are $L_E$ and $C_{V,eff}$. For clarity, we convert $C_{V,eff}$ to volumetric capacity using $C_{V,eff} / \rho_E C_M = 28$ F/mAh. In figure 7A, we plot $Q$ *versus* these parameters using typical values of $c$, $L_S$, $P_S$, $P_E$ and $D_{BL}$ (see panel). For realistic values of $\rho_E C_M$ and $L_E$, $Q$ can be ×100 below $Q_{max}$, consistent with much of the variation seen in figure 2D. We note that under appropriate circumstances, solid-state diffusion effects can reduce $Q$ even further.

Finally, we can rewrite equation 2 to represent areal capacity in terms of $Q$:

$$\frac{C}{A} = L_E \rho_E C_M \left[ 1-(RL_E^2/Q)^n \left(1-e^{-(RL_E^2/Q)^{-n}}\right)\right] \tag{14}$$

Combining equation 14 with equation 13b allows us to predict the performance of an electrode material under a range of circumstances. For example, in figure 7B we plot the areal capacity of a Si-based electrode (assuming $\sigma_E$>>1 S/m, see panel for other parameters) as a function of rate and electrode thickness. This suggests that thick Si-based electrodes can display exceptional low-rate performance but significant capacity reductions at higher rates.

In conclusion, we have developed a quantitative model to describe rate performance in battery electrodes. This combines a semi-empirical model for capacity as a function of rate with simple expressions for the diffusive, electrical and kinetic contributions to the characteristic time associated with charge/discharge. This model is completely consistent with a wide range of results from the literature and allows quantitative analysis of data by fitting to yield numerical values of parameters such as electrode conductivity and diffusion coefficients. In addition, this model can be used to predict the performance of electrode systems.

**Data Availability**: The datasets generated during and/or analysed during the current study are available from the corresponding author on reasonable request

**Acknowledgments:** All authors acknowledge the SFI-funded AMBER research centre (SFI/12/RC/2278) and Nokia for support. JNC thanks Science Foundation Ireland (SFI, 11/PI/1087) and the Graphene Flagship (grant agreement n°604391) for funding. VN thanks the European Research Council (SoG 3D2D Print) and Science Foundation Ireland (PIYRA) for funding. Dr. Ruiyuan Tian thanks Dr. Chuanfang (John) Zhang and Dr. Sebastian Barwich for useful discussions.

**Author contributions:** R.T. and S.-H.P. contributed equally to this work. R.T. collected and catalogued the literature data. R.T, P.J.K., G.C. and J.C. performed experiments. S-H. P. V.N. and J.N.C analysed the data and developed the model. J.N.C conceived the project and wrote the paper with help from R.T and S.-H.P. All authors discussed the results and commented on the manuscript. Authors declare no competing interests.

List of symbols used:

| | |
|---|---|
| $a,b,c,\beta_1$-$\beta_5$ | Compound parameters (i.e. parameters made up of combinations of other parameters) |
| $C_{SC}/M$ | Measured specific capacitance in supercapacitors |
| $C_{SC,M}$ | Intrinsic specific capacitance of a supercapacitor |
| $C_{eff}$ | Effective capacitance associated with battery electrode [F] |
| $C_{V,eff}$ | Effective volumetric capacitance associated with battery electrode [F/cm$^3$] |
| $C/M$ | Measured specific capacity for batteries |
| $C_M$ | Intrinsic specific capacity for batteries |
| $D_{BL}$ | Bulk liquid diffusion coefficient of electrolyte |
| $D_P$ | Li ion diffusion coefficient in the electrolyte-filled pores within the electrode |
| $D_S$ | Li ion diffusion coefficient in the electrolyte-filled pores within the separator |
| $D_{AM}$ | Li ion diffusion coefficient in the solid active material |
| $L_E$ | Battery electrode thickness |
| $L_S$ | Battery separator thickness |
| $L_{AM}$ | Active material thickness |
| $M_f$ | Conductive additive mass fraction |
| $n$ | Battery rate exponent |
| $P_E$ | Porosity of electrode |
| $P_S$ | Porosity of separator |
| $Q$ | Battery transport coefficient |
| $r$ | Radius of active material particles |
| $R$ | Fractional charge/discharge rate for batteries |
| $R_T$ | Charge discharge rate above which capacity begins to decay |
| $s$ | Percolation exponent |
| $t_c$ | Characteristic time associated with electrochemical reaction at the electrode/electrolyte interface |
| $\Delta V$ | CV voltage window in supercapacitors |
| $\rho_E C_M$ | Intrinsic volumetric capacity of battery electrode |
| $\sigma_E$ | Out-of-plane electrode conductivity |
| $\sigma_{BL}$ | Bulk liquid conductivity of electrolyte |
| $\sigma_M$ | Out-of-plane electrode conductivity of active material |
| $\sigma_0$ | Percolation constant |
| $\tau_{SC}$ | RC time constant in supercapacitors |
| $\tau$ | Characteristic time associated with charge/discharge for batteries |
| $\nu$ | CV scan rate in supercapacitors |

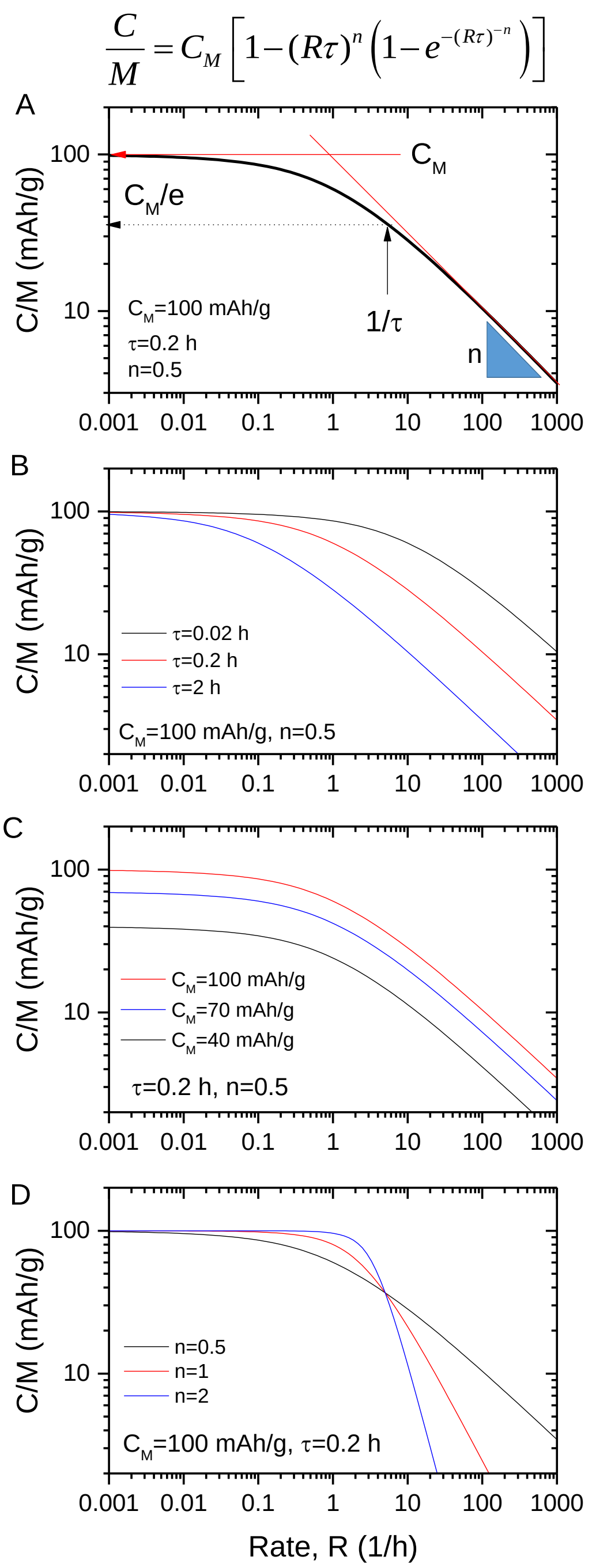

**Figure 1: Understanding the effect of the parameters defining the model.** A) Specific capacity plotted *versus* rate using equation 2 (also given above panel A) using the parameters given in the panel. The physical significance of each parameter is indicated: $C_M$ represents the low-rate limit of *C/M*, *n* is the exponent describing the fall-off of *C/M* at high rate and $\tau$ is the characteristic time. The inverse of $\tau$ represents the rate at which *C/M* has fallen by 1/e compared to its low-rate value. B-D) Plotting equation 2 while separately varying $\tau$ (B), $C_M$ (C) and n (D).

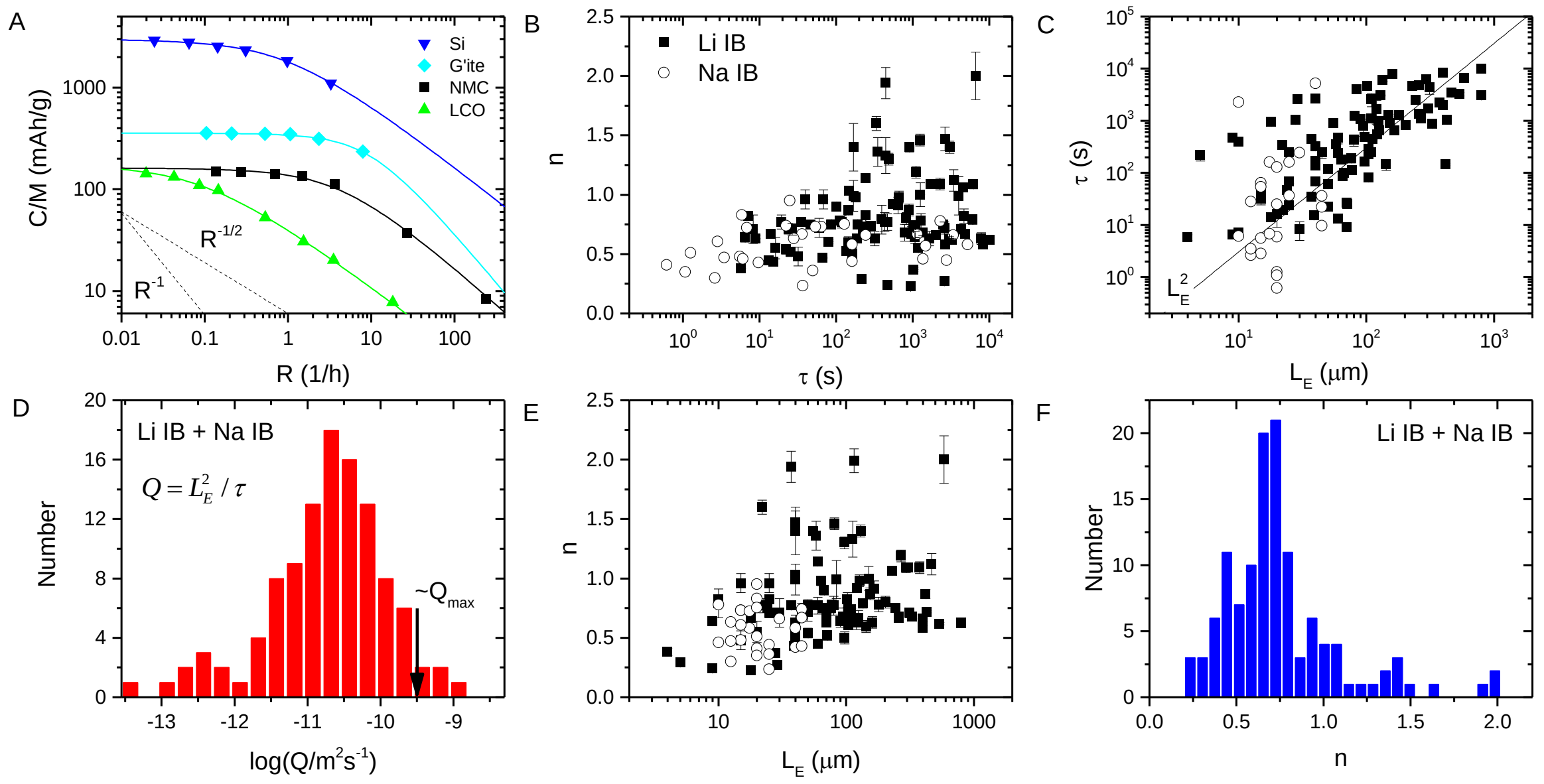

**Figure 2: Overview of literature data analysed using equation 2.** A) Four examples of specific capacity (*C/M*) *versus* rate data taken from the literature. These data all represent lithium ion half cells with examples of both cathodes and anodes. The cathode materials are nickel manganese cobalt oxide (NMC, ref[34]) and lithium cobalt oxide (LCO, ref[29]) while the anode materials are silicon (Si, ref[38]) and graphite (G'ite, ref[46]). In each case the solid lines represent fits to equation 2 while the dashed lines illustrate $R^{-1}$ and $R^{-1/2}$ behaviour. B) Equation 2 was used to analyse 122 capacity-rate data sets from 42 papers describing both lithium ion (Li IB) and sodium ion (Na IB) half cells. The resultant $n$ and $\tau$ data are plotted as a map in figure 2B (this panel does not include work which varies the content of conductive additive). C) Characteristic time, $\tau$, plotted *versus* electrode thickness, $L_E$ for Na IBs and Li IBs. The line illustrates $L_E^2$ behaviour. D) Histogram (N=122) showing frequency of occurrence of $Q = L_E^2 / \tau$ for Na IBs and Li IBs (log scale). The arrow shows the predicted maximal value of $Q$. E) Exponent, $n$, plotted *versus* electrode thickness, $L_E$, for Na IBs and Li IBs. F) Histogram (N=122) showing frequency of occurrence of $n$ for Na IBs and Li IBs.

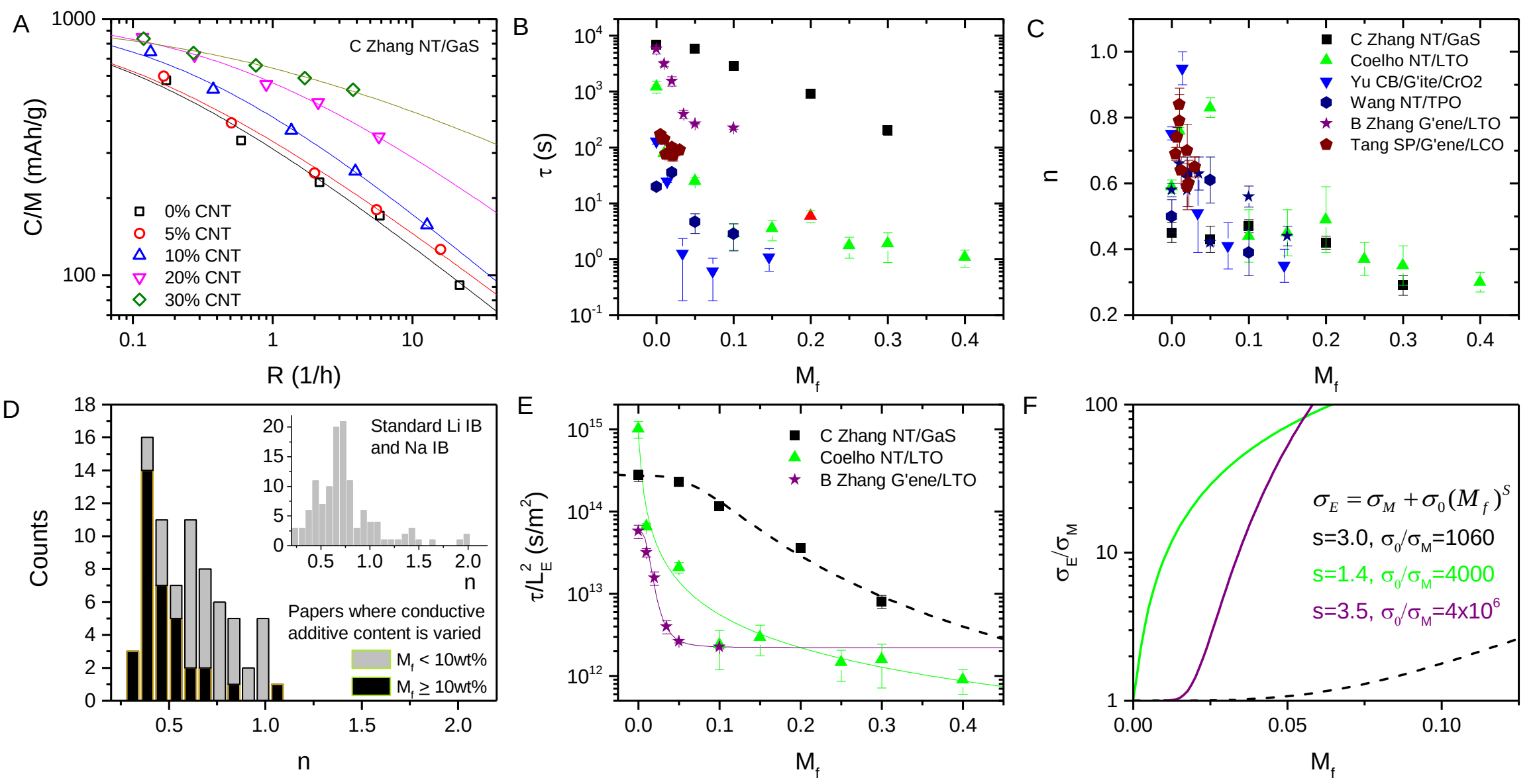

**Figure 3: The effect of varying the content of conductive additives.** A) Specific capacity *versus* rate data for lithium ion anodes based on composites of GaS nanosheets and carbon nanotubes with various nanotube mass fractions (ref[7]). The solid lines are fits to equation 2. B-C) Characteristic time (B) and exponent (C), extracted from six papers (refs[7,18,60,62-64]), plotted *versus* the mass fraction, $M_f$, of conductive additive. D) Histogram (N=75) showing frequency of occurrence of *n* in studies which varied the conductive additive content. The histogram contains data from the papers in B as well as additional refs,[19,65-68] and is divided between electrodes high and low $M_f$. The inset replots the data from 2F for comparison. E) Data for $\tau / L_E^2$ plotted *versus* $M_f$ for three selected papers.[7,18,62] The solid lines are fits to equation 6 combined with percolation theory (equation 7). F) Out of plane conductivity, $\sigma_E$, of composite electrodes normalised to the conductivity of the active material alone, $\sigma_M$. This data is extracted from the fits in 3E with the legend giving the relevant parameters. N.B. the legend/colour-coding in C applies to B, C, E, F. All errors in this figure are fitting errors combined with measurement uncertainty.

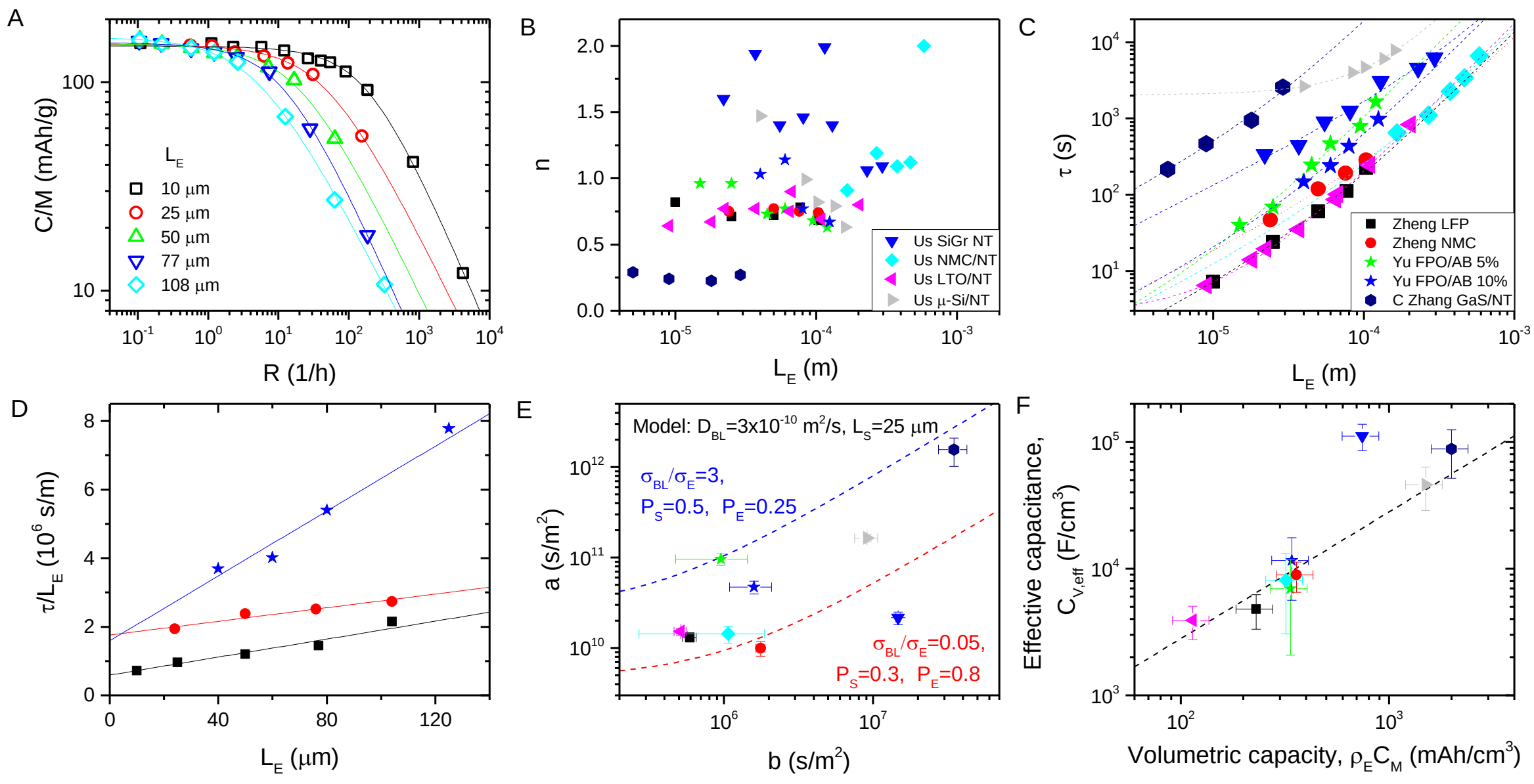

**Figure 4: The effect of varying electrode thickness.** A) Specific capacity *versus* rate data for LiFePO4-based lithium ion cathodes of different thicknesses.[17] The solid lines are fits to equation 2. B-C) Exponent (B) and characteristic time (C) plotted *versus* electrode thickness for eight data sets including four measured by us and five from the literature.[7,16,17] The legends in B and C both apply to panels B-F. The dashed lines in C) are fits to the polynomial given in equation 6. D) Plots of $\tau / L_E$ *versus* $L_E$ for a subset of the curves in C, showing the *c*-terms to be negligible (true for all data in C except the µ-Si/NT and NMC/NT data sets). E) *a*-parameter plotted *versus* *b*-parameter (see equation 6) for the data in C. The lines are plots of equation 8 using the parameters given in the panel and represent limiting cases. F) Effective volumetric capacitance, estimated from the *b*-parameters plotted *versus* the volumetric capacity, $\rho_E C_M$ . The dashed line is an empirical curve which allows $C_{V,eff}$ (F/cm$^3$) to be estimated from $\rho_E C_M$ (mAh/cm$^3$): $C_{V,eff} / \rho_E C_M = 28$ F/mAh . All errors in this figure are fitting errors combined with measurement uncertainty.

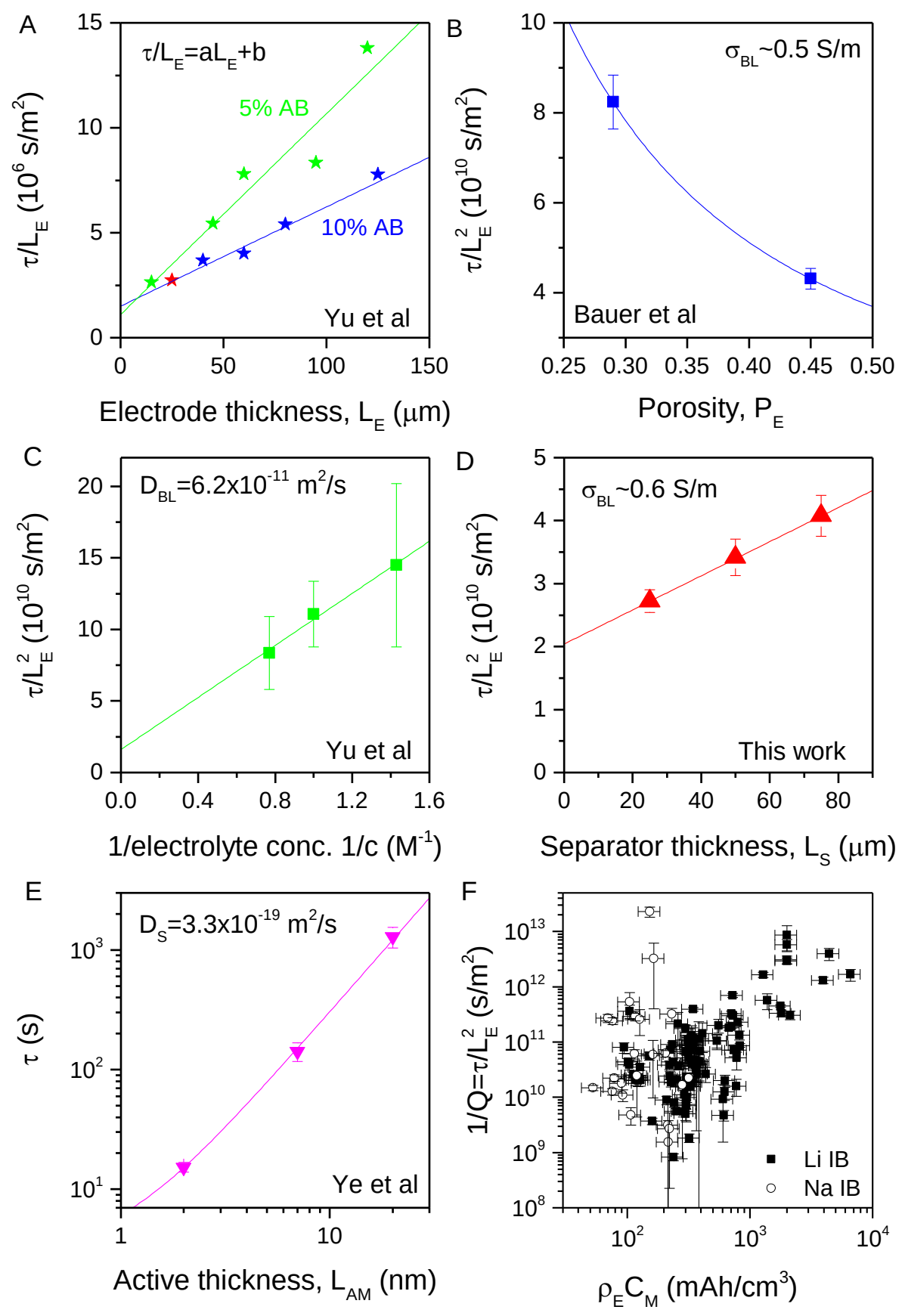

**Figure 5: Further testing of the terms in equation 6**. A) $\tau / L_E$ *versus* $L_E$ for electrodes with 5 and 10% acetylene black, and so different conductivities (extracted from ref[16]). This results in different *a*-parameters (slopes) but the same *b*-parameter (intercept), consistent with equation 6. B) $\tau / L_E^2$ *versus* porosity extracted from ref[19]. The line is a fit to equation 9 and yields a value of $\sigma_{BL}$ close to the expected value (see panel). C) $\tau / L_E^2$ *versus* inverse electrolyte concentration extracted from ref[16]. The line is a fit to equation 10 and yields $D_{BL}$ close to the expected value (see panel). D) $\tau / L_E^2$ *versus* separator thickness (this work). The line is a fit to equation 11 and yields $\sigma_{BL}$ close to the expected value (see panel). E) Characteristic time *versus* the thickness of a thin active layer ($TiO_2$) extracted from ref[12]. The line is a fit to equation 12 and yields a diffusion coefficient for Li ions in anatase $TiO_2$ close to the expected value.[81] F) $1/Q$ plotted *versus* the intrinsic volumetric electrode capacity, $\rho_E C_M$, for cohorts I and II showing the scaling predicted by equation 13a. All errors in this figure are fitting errors combined with measurement uncertainty.

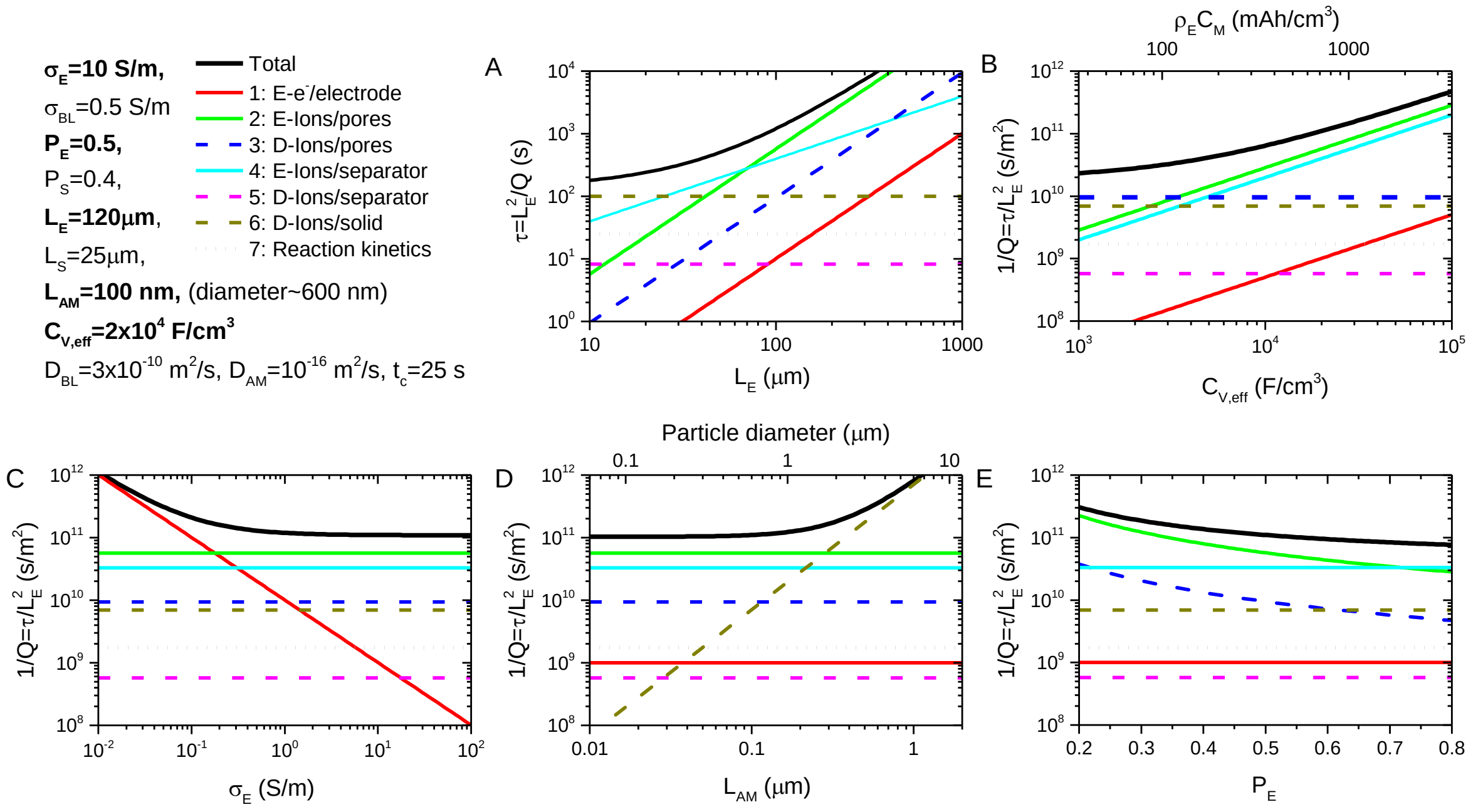

**Figure 6: Comparison of the magnitude of terms 1-7 in equation 13a, as well as their sum, for a range of electrode parameters.** Note that, while in A, $\tau$ is plotted *versus* $L_E$, in all other panels, $Q^{-1}$ is plotted *versus* the relevant parameter. The parameters used are given at the top left. These values are justified in the SI. Those bold parameters were kept constant in all panels except one, where they were varied. The solid black lines represent the total value of $\tau$ or $Q^{-1}$. Low values of both $\tau$ and $Q^{-1}$ are needed for good rate performance. The other curves represent the seven individual terms in equation 13a, labelled 1-7 (numbered from left to right in the equation). Electrical and diffusion limited terms are marked as solid and dashed lines respectively with the reaction kinetics term represented by grey dots. The legend in the top left gives the term number as well of a summary of what it represents. Those terms labelled by "E" are electrically limited while those labelled by "D" are diffusion limited. The top axis in figure 6B represents the volumetric capacity of the electrode calculated using $C_{V,eff} / \rho_E C_M = 28\ \text{F/mAh}$. N.B., $L_{AM}$=100 nm corresponds to a particle diameter of $2r\approx600$ nm because $L_{AM}=r/3$ for pseudo-spherical particles.

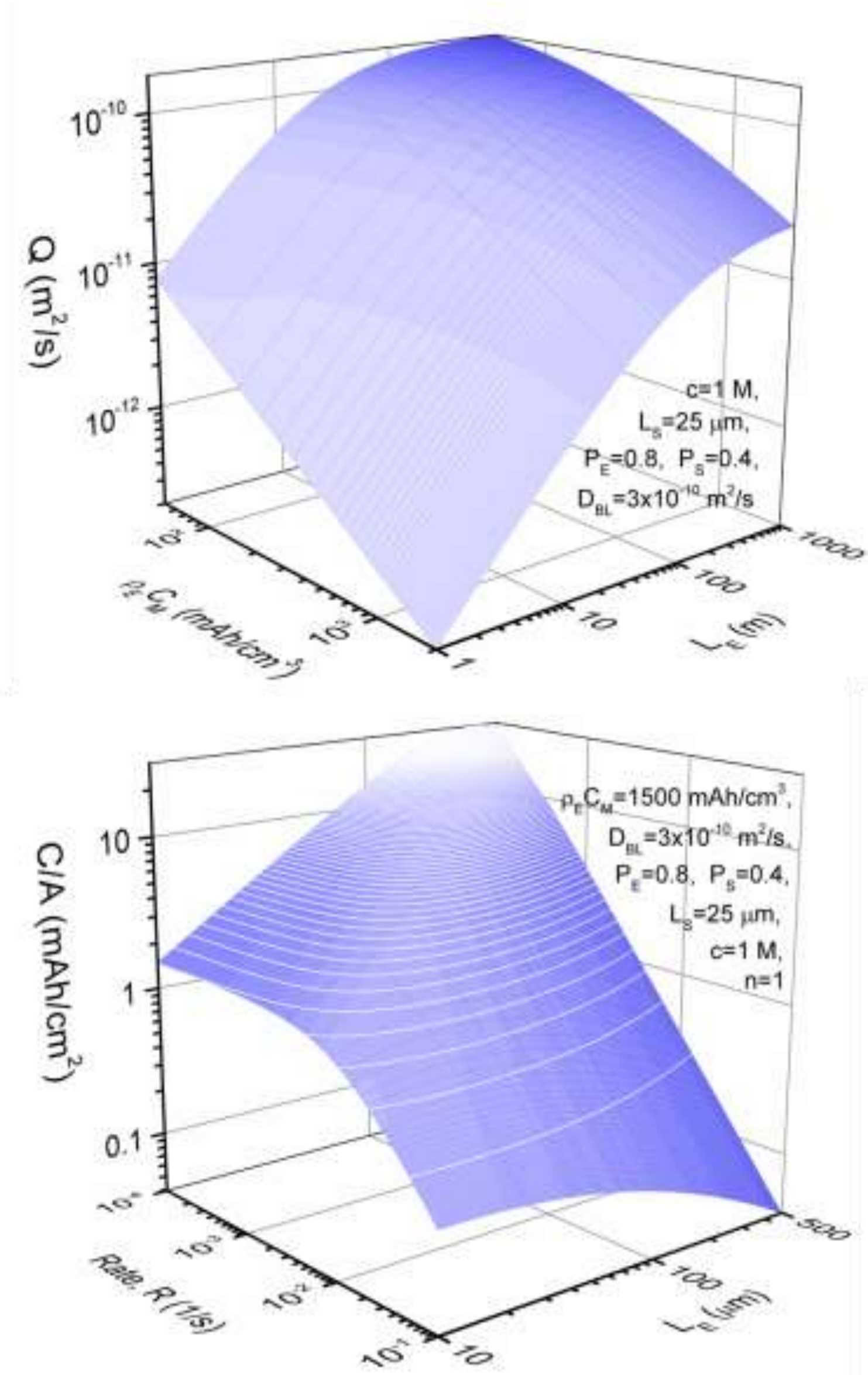

**Figure 7: Predictive analysis.** A) Transport coefficient, *Q*, plotted *versus* volumetric capacity and electrode thickness using equation 13b. In A, the effective volumetric capacitance is converted to volumetric capacity using $C_{V,eff} / \rho_E C_M = 28$ F/mAh. B) Areal capacity for Si-based electrodes (defined by $\rho_E C_M$ =1500 F/cm$^3$) plotted *versus* rate and electrode thickness. All parameters used in the calculations are given in the panels. N.B. we take *n*=1 as figure 6 implies thick electrodes to be electrically limited.